\documentclass[aps,pre,superscriptaddress,showkeys,longbibliography,twocolumn]{revtex4-2}

\usepackage{amsmath}
\usepackage{graphicx}
\usepackage{float,physics}
\usepackage{siunitx}
\usepackage{xcolor}
\usepackage{epstopdf}
\definecolor{darkgreen}{rgb}{0,0.6,0.0}

\usepackage{latexsym}
\usepackage{amsmath}
\usepackage{amsfonts}
\usepackage{amssymb}
\usepackage{mathrsfs}
\usepackage{verbatim}
\usepackage{anysize}
\usepackage{soul}

\usepackage[utf8]{inputenc}
\usepackage{algorithmic}
\usepackage{algorithm}
\usepackage{enumerate}
\usepackage{amssymb, amsmath, amsthm}
\usepackage{graphicx}
\usepackage{lmodern,url}
\usepackage{graphicx}
\usepackage{wrapfig}
\usepackage{hyperref}

\floatname{algorithm}{Algoritmo}

\newcommand{\bo}{\raise-1mm\hbox{\Large$\Box$}}

\usepackage{multirow}

\usepackage{hyperref}
\hypersetup{
    colorlinks=true,
    linkcolor=blue,    % Color of internal links
    citecolor=blue,    % Color of citation links
    urlcolor=blue       % Color of external links
}

\def\vv{{\bf v}}
\def\vee{\hat{\bf e}}

\def\vr{{\bf r}}
\def\beq{\begin{equation}}
\def\eeq{\end{equation}}

\def\Pe{\mathrm{Pe}}

\begin{document}
\title{Role of Translational Noise in Motility-Induced Phase Separation of Hard Active Particles}
\normalsize
\author{Felipe Hawthorne}
\email[]{felipehawthorne@ufpr.br}
%\email[]{jfreire@fisica.ufpr.br }
\affiliation{Departamento de Física, Universidade Federal do Paraná, 81531-990, Curitiba-PR, Brazil}
\author{Pablo de Castro}
%\email[]{pablo.castro@ictp-saifr.org}
\affiliation{ICTP-South American Institute for Fundamental Research - Instituto de F\'isica Te\'orica da UNESP, Rua Dr.~Bento Teobaldo Ferraz 271, 01140-070 S\~ao Paulo, Brazil.}
\author{José A.\ Freire}
\email[]{jfreire@fisica.ufpr.br}
\affiliation{Departamento de Física, Universidade Federal do Paraná, 81531-990, Curitiba-PR, Brazil}
\begin{abstract}
Self-propelled particles, like motile cells and artificial colloids, can spontaneously form macroscopic clusters. This phenomenon is called motility-induced phase separation (MIPS) and occurs even without attractive forces, provided that the self-propulsion direction fluctuates slowly. In addition to rotational noise, these particles may experience translational noise, not coupled to rotational noise, due to environmental fluctuations. We study the role of translational noise in the clustering of active Brownian hard disks. To tease apart the contribution of translational noise, we model excluded-volume interactions through a Monte-Carlo-like overlap rejection approach. We find that increasing translational diffusivity has a non-monotonic effect on clustering. At low values, it makes clusters more compact and rounded (less filamentous), eventually promoting genuine MIPS. For sufficiently higher translational diffusivity, clusters evaporate. We develop a theory for the cluster mass distribution, and employ a hydrodynamic approach with parameters taken from the simulation, that explains the clustering phase diagram.

\end{abstract}
\maketitle

\section{Introduction}
\label{introduction}
Active particles, such as motile organisms and artificial colloids, can consume local fuel to self-propel~\cite{bechinger2016active,Fier18,Met15}. Typically, their self-propulsion direction fluctuates stochastically and slowly, as observed in run-and-tumble~\cite{villa2020run,garcia2021run,figueroa20203d}, active Brownian~\cite{rojas2023wetting,buttinoni2012active} and colloidal \cite{PhysRevLett.110.238301, Redner13} particles. As a result of their persistent self-propulsion, active particles are able to cluster together even in the absence of attractive forces. 

During the self-propulsion persistence timescale, particles can block each other and thus trigger an exponential growth of clusters which eventually generates a stationary distribution of cluster masses~\cite{Su23,Omar23,Schmi23}. For sufficiently persistent and dense systems, this process results in a true phase separation phenomenon known as motility-induced phase separation (MIPS)~\cite{Tai08,Cates15, Marchetti12,fily2014freezing,zhao23,stenh21,Su23}, in which macroscopic clusters of active particles spontaneously emerge. The complete phase diagram describing clustering behavior in active particle systems can be considerably more intricate. For instance, soft disks can exhibit liquid, hexatic, and solid phases~\cite{10.1103/PhysRevLett.121.098003,10.1103/PhysRevLett.125.178004,10.1103/PhysRevResearch.2.023010,San22} and particle softness itself can substantially modify the conditions under which phase separation occurs~\cite{PhysRevE.103.052605}. Additionally, colloidal suspensions featuring explicit passive attraction display non-monotonic behavior as self-propulsion persistence increases, with clusters first being promoted and then suppressed~\cite{10.1103/PhysRevE.88.012305}. Similar non-monotonic effects are also reported in systems of repulsive active dumbbells~\cite{10.1209/0295-5075/108/56004,10.1039/d3sm01030a}.

There is strong experimental evidence that motile cells can be modeled as persistent walkers subject to an additional source of noise of the translational kind, which in turn typically comes from thermal fluctuations in the surrounding environment~\cite{Wu14,Met15,Fier18,Schi93,Abdul22}. However, the role of translational noise in the clustering behavior of active particles has been somewhat overlooked. When active particles interact via hard potentials and there is no translational noise, clusters are gel- or fractal-like as opposed to more rounded. In fact, using a kinetic Monte Carlo model of self-propelled \emph{hard disks}, it was shown that translational noise is essential
for the formation of large clusters~\cite{Ber14}. When soft potentials are used~\cite{Marchetti12,Redner13}, one observes large clusters even in the absence of translational noise. As pointed out in Ref.~\cite{Ber14}, soft potentials add a displacement non-collinear to the persistent displacement, something that a hard potential cannot provide since it simply forbids persistent displacements that lead to particle overlap. Therefore, for hard particles, the only source of displacement non-collinear to the persistent direction is the translational noise. As we shall see, this perception is in agreement with our observations discussed below that a very small translational diffusivity greatly changes the phase diagram of the clustering behavior.

In this work, we quantify the role of the translational noise in the clustering behavior of a system of 2D active Brownian hard disks interacting through overlap rejection. To do that, we study cluster masses and morphologies in the presence of translational diffusivity.

Specifically, we measure the fraction of particles within the largest cluster ($f_{\rm max}$) and demonstrate that this fraction exhibits a non-monotonic dependence on translational diffusivity. We introduce a master equation model~\cite{Ber14} to describe the time evolution of the average number of clusters of each size, i.e., the cluster size distribution (CSD), which provides insights into the numerically obtained CSD data. Additionally, we develop a method to extract the density-dependent velocity directly from the simulations, which is essential for the hydrodynamic description of MIPS~\cite{Bialk__2013, Cates15}. Using this velocity function, we trace a rough estimate of the phase-separation local instability line, i.e., the spinodal line, within the phase diagram of our system, parameterized by disk density, self-propulsion speed, and translational noise.

This paper is organized as follows: Section~\ref{model} outlines the model used in this study. In Section~\ref{results}, we present the numerical results from our simulations, including the order parameter curves and phase diagram, alongside theoretical findings derived from the aggregation master equation model and the mean-field hydrodynamic model based on local density. Finally, Section~\ref{conclusion} summarizes our main conclusions.

\section{The Model}
\label{model}
\subsection{The individual disk dynamics}
\label{individual}
We employ the active Brownian particle (ABP) model in two dimensions with added 
translational noise. 
Unlike typical models for artificial colloids where rotational and translational 
diffusivities are coupled~\cite{Redner13}, we treat these diffusivities as independent 
parameters, reflecting the assumption that they originate from distinct noise sources.

Accordingly, the internal angular variable of the disk, $\theta$, and its center 
position $\mathbf{r}$, evolve over time according to the following overdamped Langevin equations:
\beq
\begin{split}\label{Langevin}
\frac{d\vr}{dt} =&\, v_0 \,\vee_\theta + \sqrt{2 D_T}\,\boldsymbol{\xi}_\vr, \\
\frac{d\theta}{dt} =&\, \sqrt{2D_{\textrm{R}}} \,\xi_\theta, 
\end{split}
\eeq
where $\vee_\theta=(\cos\theta, \sin\theta)$ is the unit vector along the self-propulsion direction of the disk and the three noise terms, $\xi_x, \xi_y$ and $\xi_{\theta}$ are independent Gaussian white noises,
$$\langle \xi_k(t)\rangle=0,$$ 
$$\langle \xi_k(t)\xi_l(t')\rangle = \delta_{kl}\delta(t-t'), \quad k,\,l = \{\theta,x,y\}.$$
The $v_0$ term in the first equation, together with the equation for $\theta$, define a persistent random walk (RW) with a persistence time $\tau\equiv (2 D_{\textrm{R}})^{-1}$. The term proportional to $\sqrt{D_T}$, the translational noise, adds a purely diffusive displacement to the persistent motion. A brief discussion of the mean-squared displacement (MSD) arising from these dynamics is provided in Appendix~\ref{app:msd}. 

From Eq.~\eqref{Langevin}, which describes the stochastic evolution of a single particle, we directly derive the Fokker–Planck equation for the single-particle probability distribution function, $\psi(\vr,\theta,t)$, presented here for future reference:
\beq\label{FP}
\partial_t \psi = -\nabla_\vr \cdot [v_0 \,\vee_\theta\,\psi - D_T \nabla_\vr \psi] + \frac{1}{2\tau} \partial_{\theta\theta} \psi.
\eeq

\subsection{The collective disk dynamics}
\label{collective}
We now consider a system of $N$ hard disks of radius $a$ following the dynamics of Eq.~\eqref{Langevin} in an area of size $L^2$ with periodic boundary conditions. The only interaction between the disks is overlap rejection, i.e., the impediment of superposition. Together with $v_0$ and $D_T$, the packing fraction $\phi=N\pi a^2/L^2$ completes the set of parameters that characterizes the system. 

Our results will be presented using the radius of the disks $a$ as the unit of length and the persistent time $\tau$ as the unit of time. In these units, the self-propulsion speed is equal to the Péclet number, ${\rm Pe}=v_0\tau/a$, defined purely in terms of the rotational diffusivity, as in Refs.~\cite{Fily2012,Moore2023,Ber14}. Similarly, the translational diffusion coefficient $D_T$ will be presented in units of $a^2/\tau$.

At the start of the simulation, the square of side length $L$ is populated with $N$ non-overlapping disks, randomly distributed such that $N$ matches the desired packing fraction, $\phi$.
At each step of the simulation, the positions and self-propelled directions of all disks were simultaneously updated according to a discrete version of Eq.~\eqref{Langevin} through a forward Euler-Maruyama scheme,
\beq
\begin{split}\label{LangevinEM}
\vr_{n+1} =& \,\vr_n + v_0\,\vee_\theta \, dt + \sqrt{2D_Tdt} \, {\cal N}_{n}^{\vr}, \\
    \theta_{n+1} =& \,\theta_n + \sqrt{2D_Rdt}\,{\cal N}^{\theta}_{n},
\end{split}
\eeq
where ${\cal N}^{\{\theta,x,y\}}_n$ are three independent Gaussian variables with zero mean and unit variance. 

To enforce hard-core repulsion, any disks that overlapped after being moved were repositioned to their previous positions simultaneously. This process could lead to newly occurring overlaps of the repositioned disks. Thus, the procedure of identifying and eliminating the overlapping disks was repeated until none was identified. Upon reaching this condition, the time step was considered complete.

When analyzing clustering in systems of impenetrable disks such as ours, a neighborhood criterion must be defined. We considered any two disks to belong to the same cluster if the center-to-center distance between them was less than or equal to $2.4$ times the radius of the disk. All snapshots, $f_{\rm max}$ curves, and CSDs were obtained for $t \geq 500\tau$. Pe was changed by varying $v_0$.

\section{Results}
\label{results}
\subsection{The impact of the translational noise on clustering}
\label{numerical}

Levis and Berthier~\cite{Ber14} found that, in hard disk systems, such as the one studied here, translational noise plays a crucial role in shaping cluster morphology. Without translational noise, i.e., when the individual disk dynamics is a simple persistent RW, the clusters form a highly dispersed, filamentous structure. However, introducing translational noise results in the emergence of more compact clusters. According to their analysis, modest translational diffusion provides the only source of displacement that is non-collinear to the persistent motion direction, thereby influencing the cluster formation process. Notably, their study did not present a systematic investigation of the role of translational noise, particularly in the large $D_T$ region. Our findings reveal, as expected, that as translational noise continues to increase, clustering is eventually disrupted. This disruption can be interpreted as the ``evaporation" of clusters due to the enhanced ``thermal" motion of their constituent particles: while a small amount of noise facilitates the clustering of hard disks, excessive noise suppresses it. 

\begin{figure}[htb]
\centering
\includegraphics[width=0.45\textwidth]{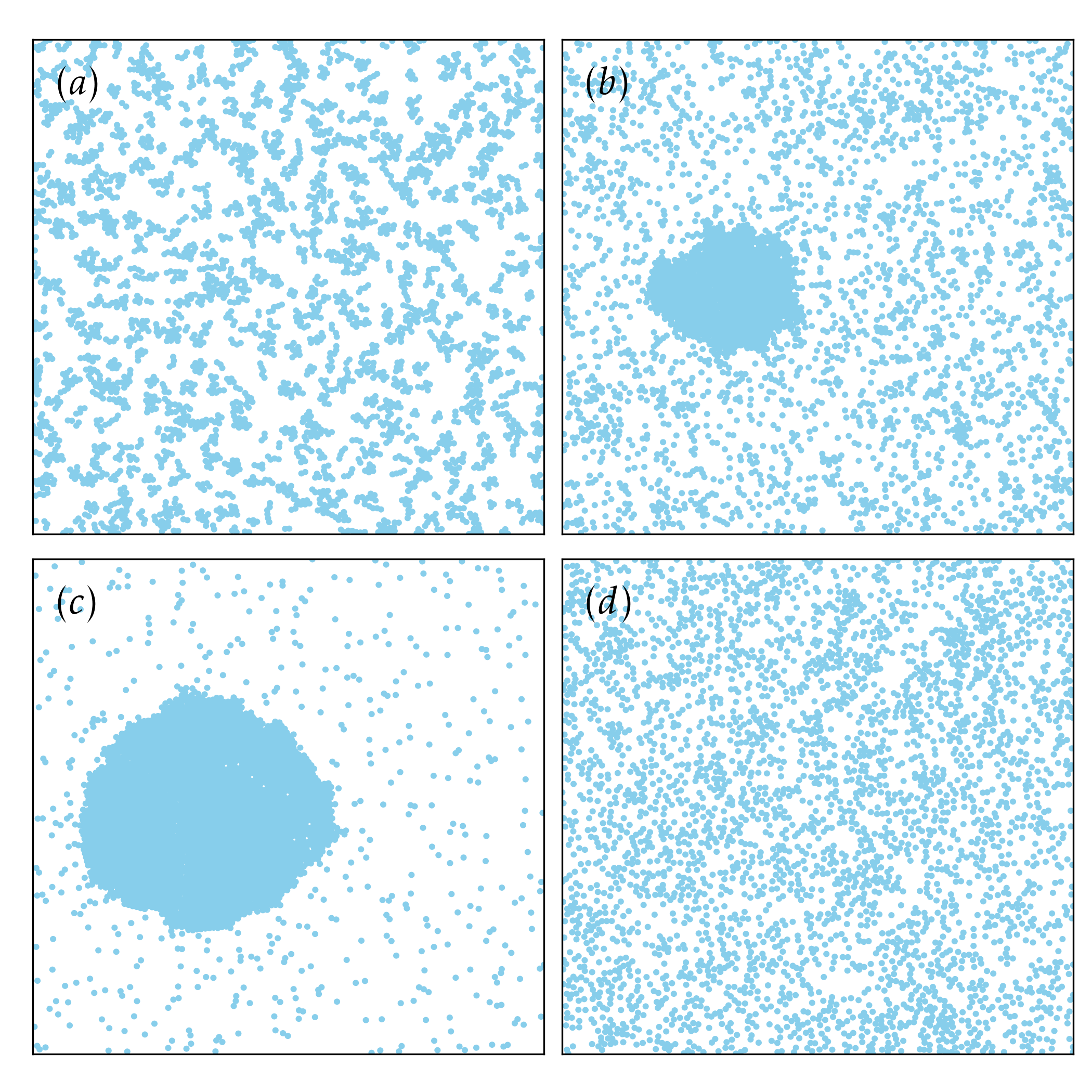}
\caption{\label{fig:Snapshots A&B}Snapshots for a system of hard disks at ${\phi = 0.178}$, $N=1755$ and $\text{Pe} = 300$, shown for four different values of the translational diffusion coefficient. From (a) to (d), $D_T \in \{0,\,7,\,14,\,200\}~[a^2/\tau]$. A moderate value of $D_T$ transforms the  filamentous structure in (a) into more compact and rounded clusters seen in (b) and (c). However, above a certain threshold, the thermal-like agitation introduced by translational diffusion fully disrupts the clustering; compare panels (c) and (d).}
\end{figure}

This is illustrated in Fig.~\ref{fig:Snapshots A&B}, which shows snapshots taken at fixed values of $\phi$ and Pe, for four different levels of translational noise $D_T$. The snapshots indicate that, at small $D_T$, the clusters are somewhat filamentous, exhibiting a certain degree of fractality. This is progressively disrupted as translational noise increases, leading to more round clusters. A detailed fractal analysis is beyond our scope, but would consist of calculating the scaling exponents relating the radius of gyration of the cluster to its mass, as done, for instance, in Refs.~\cite{PhysRevLett.131.068201,10.1103/PhysRevLett.125.178004}.
\begin{figure}
\includegraphics[width=0.4\textwidth]{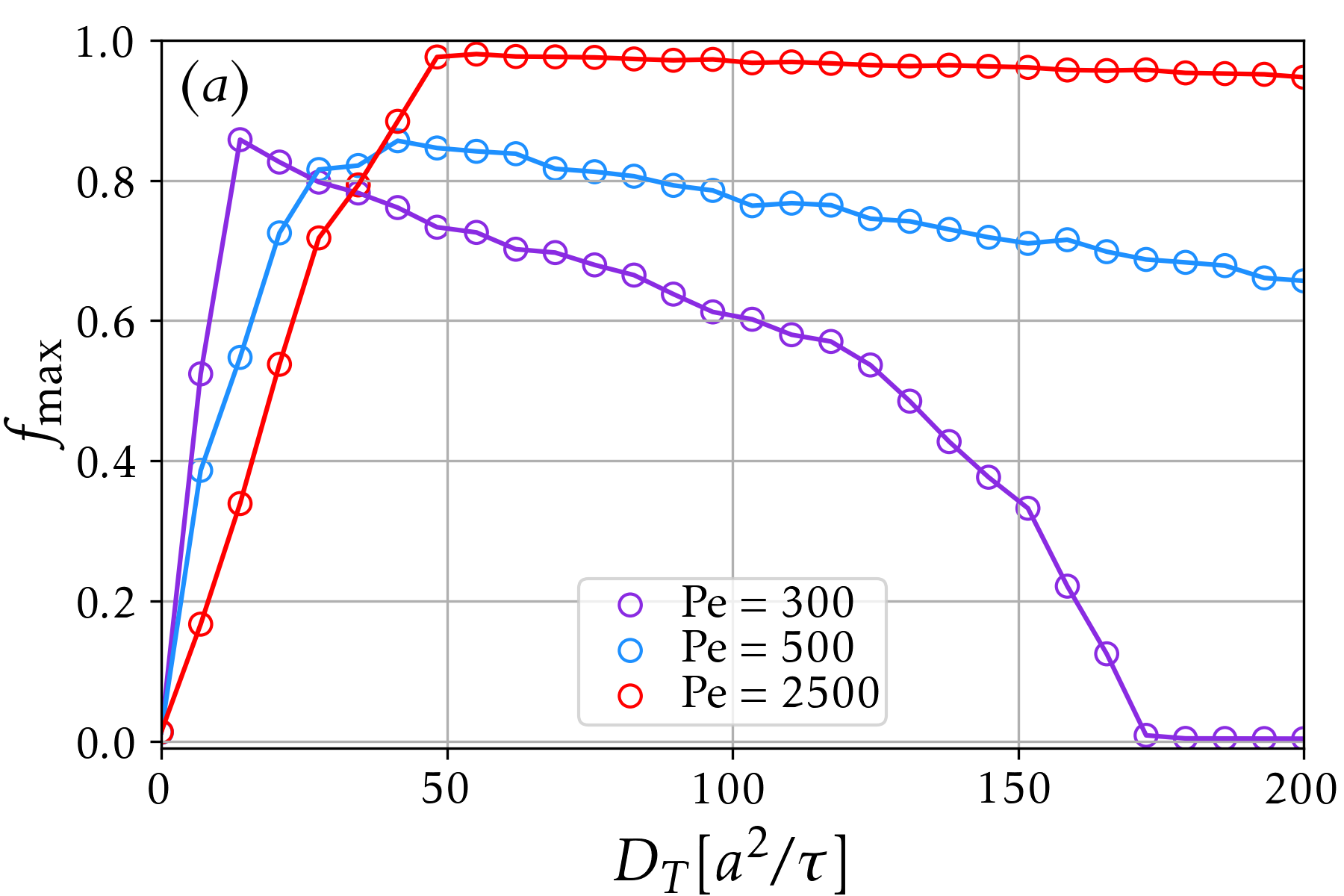}
\includegraphics[width=0.4\textwidth]{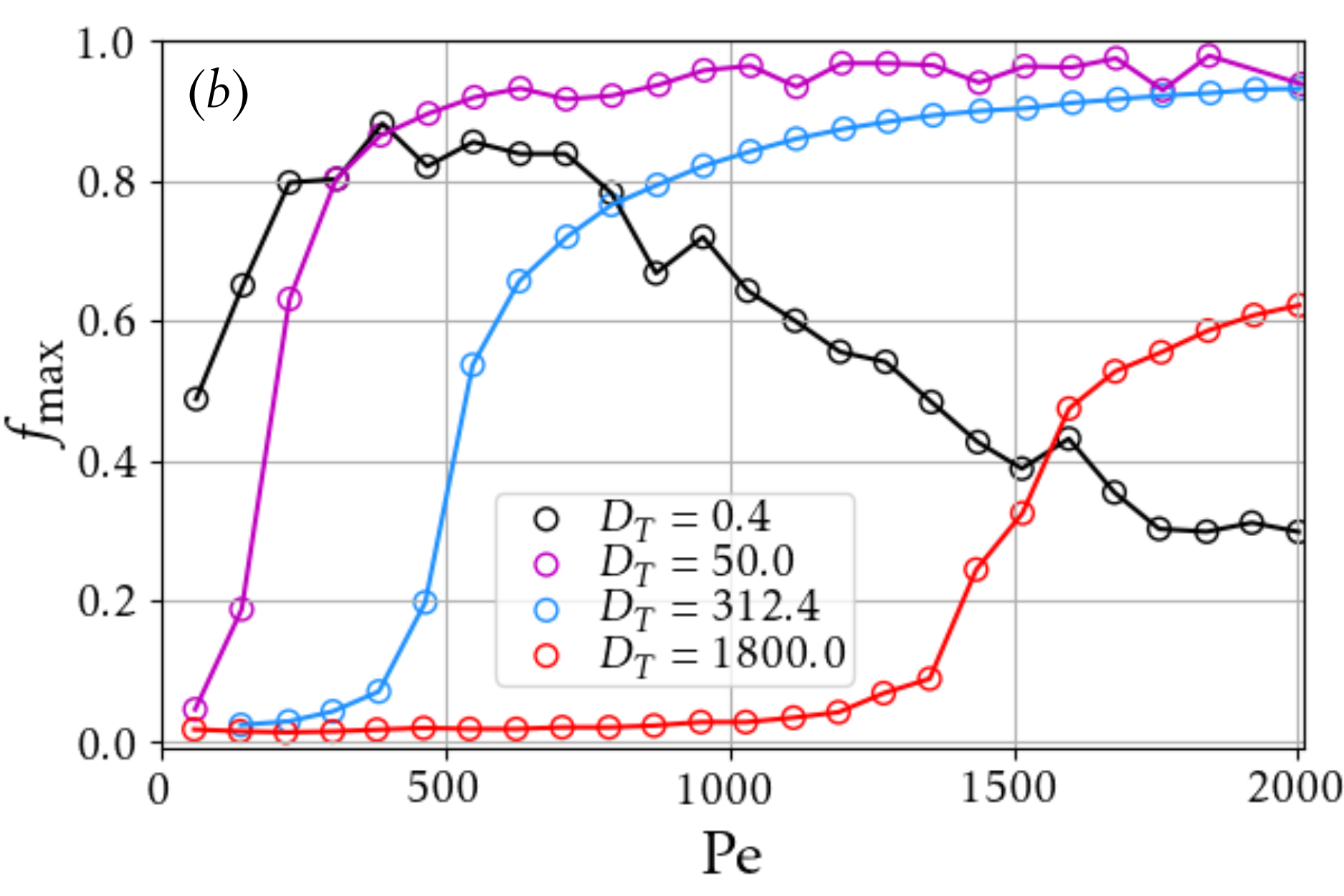}
\caption{\label{fig:Fig_Resultado1}$(a)$ The rise and fall of $f_{\rm max}$ with increasing translational noise. The systems considered have $\phi=0.178$, and three different values of Pe. $(b)$ $f_{\rm max}$ as a function of Pe for different values of $D_T$ (in units of $a^2/\tau$) and $\phi=0.4$. A larger value of $D_T$ requires a larger self-propulsion speed for the onset of clustering. In both panels, $N = 7830$. }
\end{figure}

This effect of translational noise is also reflected in $f_{\rm max}$, the fraction of disks in the largest cluster. This quantity exhibits a non-monotonic dependence on $D_T$ for fixed values of $\phi$ and Pe; see Fig.~\ref{fig:Fig_Resultado1}(a). The translational noise also has the effect of requiring a larger self-propulsion speed for the onset of clustering. Fig.~\ref{fig:Fig_Resultado1}(b) illustrates this effect with a plot of $f_{\rm max}$ versus Pe, for different values of $D_T$.

\begin{figure}[htb]
\vspace{-0.5cm}
\hspace{-0.3cm}
\includegraphics[width=0.45\textwidth]{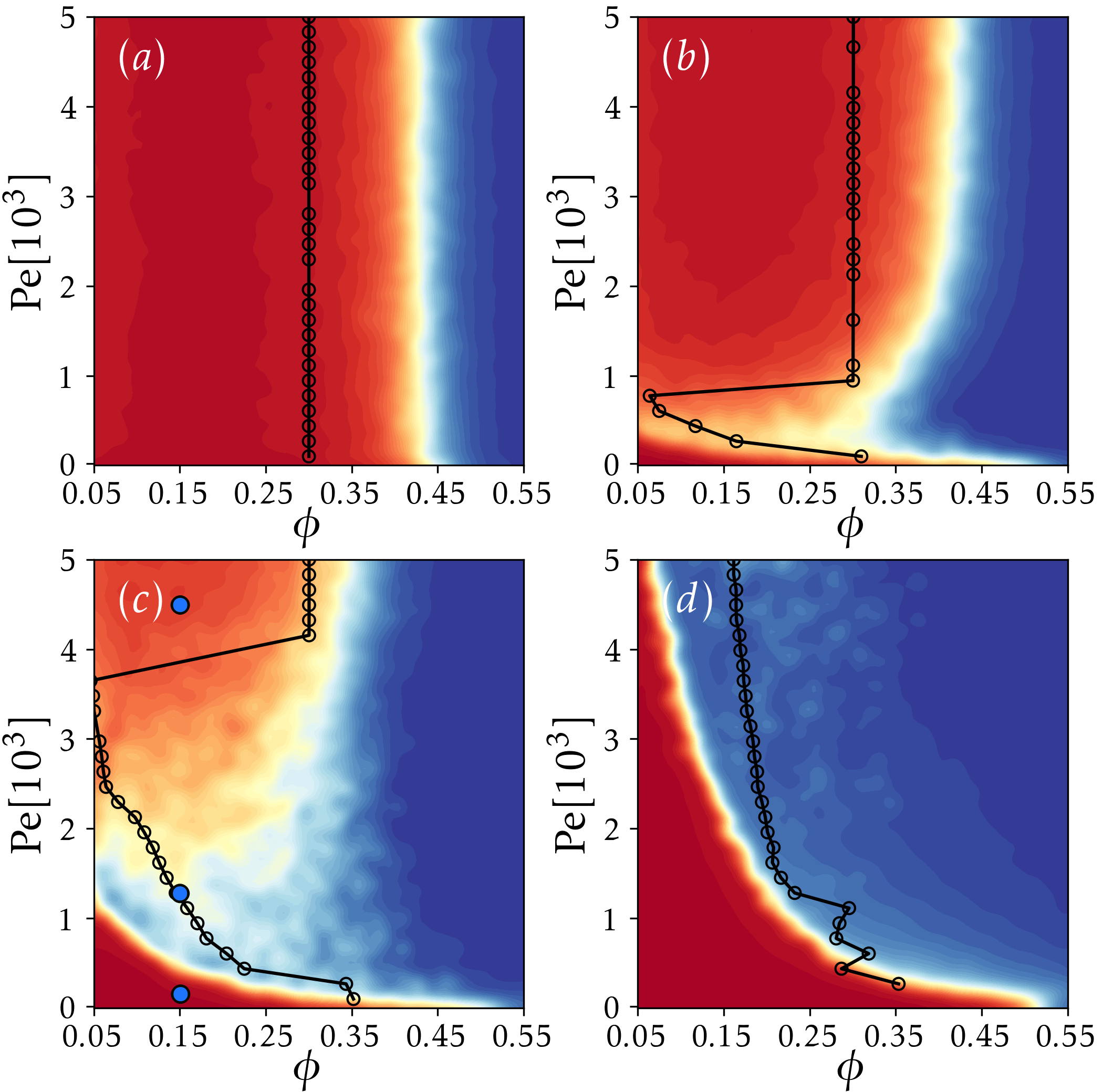}

\hspace{-0.3cm}
\includegraphics[width=0.45\textwidth]{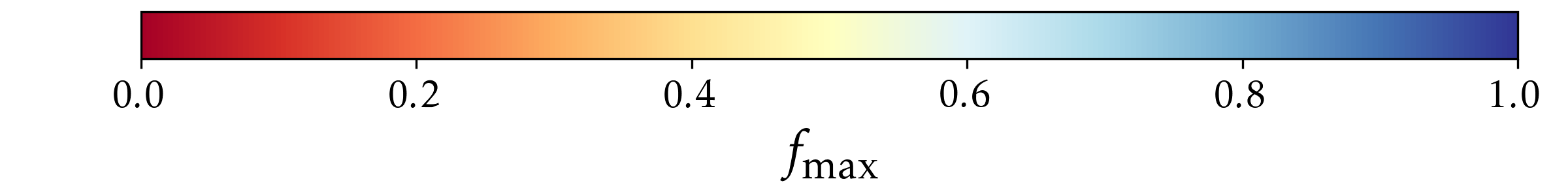}
\vspace{-0.2cm}
\hspace{-0.4cm}
\includegraphics[width=0.55\textwidth]{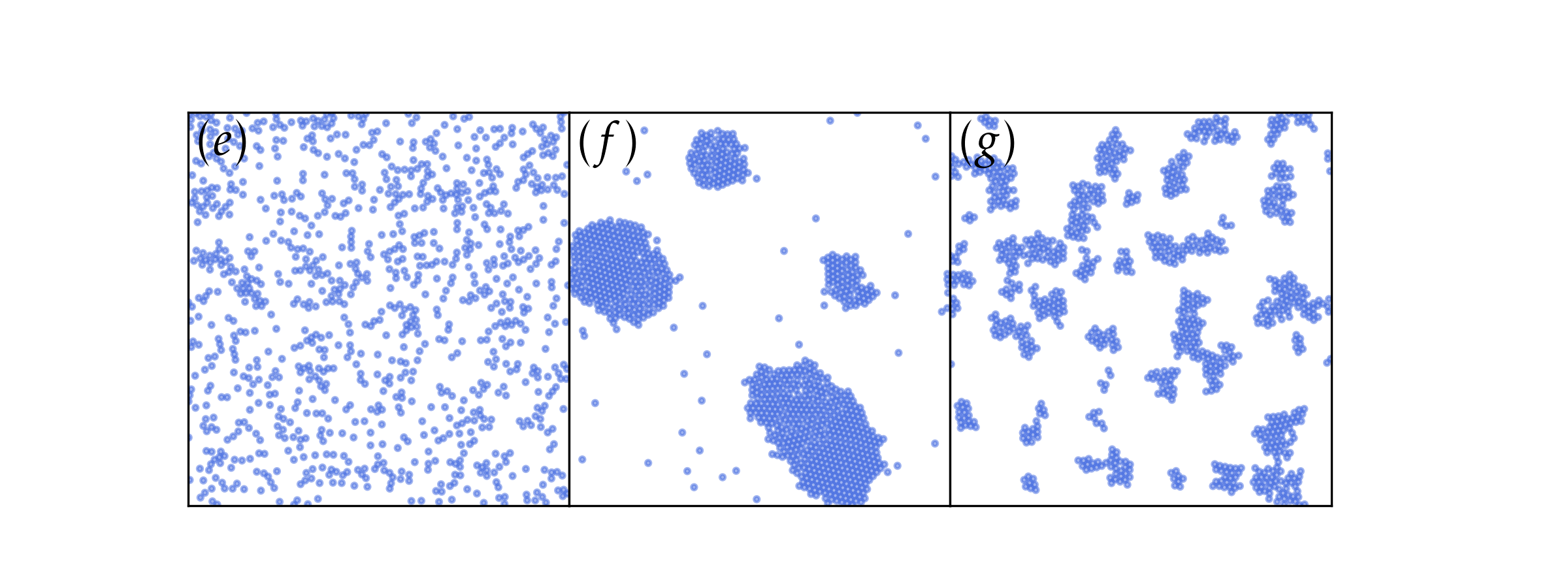}
\caption{\label{fig:PhsDiag}The fraction of disks in the largest cluster, $f_{\rm max}$, as a function of Pe and $\phi$. From (a) to (d), $D_T = \{0,\,0.5,\, 8,\, 20\}\,[a^2/\tau]$ and $L = 128\,[a]$. Snapshots for $D_T = 8\,[a^2/\tau]$, $\phi = 0.15$, $N = 1000$, and (e) ${\rm Pe} = 10^2$, (f) ${\rm Pe} = 1.3\times 10^3$, and (g) ${\rm Pe} = 4.6\times 10^3$ corresponding to the blue dots in panel (c), illustrate the disk morphology change upon increasing Pe. The black symbols and lines mark rough estimates of the spinodal lines (the clustering local instability line) obtained from the numerical simulations in Section~\ref{Dens}, with $N = 1500$ and $\phi = 0.196$.}
\end{figure}

In Fig.~\ref{fig:PhsDiag} we provide an overview of the phase diagram by plotting $f_{\rm max}$ in the plane (Pe, $\phi$) for four values of $D_T$. Another noteworthy effect is the non-monotonic dependence of clustering on the self-propulsion speed, characterized by the Péclet number; for snapshots, see Figs.~\ref{fig:PhsDiag}(e-g). This phenomenon, previously reported for disks interacting via soft potentials~\cite{Su23} and colloidal particles with attraction~\cite{10.1103/PhysRevE.88.012305}, is now observed in systems of impenetrable disks. 
In the high persistence limit ($\mathrm{Pe} \gg \sqrt{D_T\tau/a^2}$), particle motion gets dominated by self-propulsion and therefore the effect of translational noise in compacting clusters is reduced, resulting in the formation of smaller and more filamentous clusters again; see Fig.~\ref{fig:PhsDiag}(g). This results in a decrease in $f_{\text{max}}$ with increasing Pe, as seen in Fig.~\ref{fig:Fig_Resultado1}(b) (black curve) and in Figs.~\ref{fig:PhsDiag}(b) and \ref{fig:PhsDiag}(c). This behavior echoes the mechanism discussed in Ref.~\cite{10.1103/PhysRevResearch.2.023010}, where frictional interactions between soft disks play a key role in determining the phase behavior.

\subsection{The Cluster Size Distribution (CSD)}
\label{csd_result}
In the previous section, we introduced the fraction of disks in the largest cluster, $f_{\rm max}$, 
as an order parameter to characterize clustering. 
While $f_{\rm max}$ provides useful insight, a more comprehensive description of the clustering phenomenon 
is offered by the Cluster Size Distribution (CSD) \cite{de2021active,de2021diversity}, which quantifies the number of clusters with a given number of disks. 
As we shall demonstrate, the CSD exhibits a marked transition in behavior at specific critical values of the parameters: $\phi$ (packing fraction), Pe (Péclet number, representing the dimensionless self-propulsion speed), and 
$D_T\,a^2/\tau$ (the dimensionless diffusion coefficient). This transition in the CSD marks the \emph{global} instability of the micro-clustering state (where no macroscopic cluster is present), characterized by the nucleation and subsequent growth of clusters that exceed a critical size~\cite{annurev:/content/journals/10.1146/annurev-conmatphys-070909-104101,Redner13}. The \emph{local} instability will be addressed separately in Section \ref{Dens}.

To aid in the interpretation of the cluster size distributions (CSDs) obtained from the numerical experiments discussed below, we introduce a master-equation-based description of the aggregation process. This approach, akin to those employed in Refs.~\cite{Peruani_2006,Ber14}, is commonly referred to as Smoluchowski coagulation theory \cite{Smo1916}. The object of the equations is the number of $k$-clusters (clusters containing $k$ disks) at time $t$, denoted $n_k(t)$.  The underlying hypotheses behind our aggregation model are the following:
\begin{enumerate}
    \item clusters increase/decrease in size by the attachment/detachment of a single disk;
   \item the number of disks attached to a $k$-cluster during time $\tau$ is proportional to the product of the monomer density, approximated as $n_1/L^2$, and an area around the cluster of size $v_0\tau a S_k$, where $aS_k$ is the effective perimeter of a $k$-cluster and $v_0\tau$ is the persistence length;
    \item the number of disks detached from a $k$-cluster during time $\tau$ is proportional to the same $S_k$ factor.
\end{enumerate}
The effective perimeter $S_k$ will be assumed to have a power-law dependence on the number of disks in the cluster, $S_k \propto k^\omega$ (with the overall constant being absorbed in the constants $c_a$ and $c_d$ below). Given hypothesis number 1 above, it would be natural to expect that $S_k$ should be proportional to the actual number of disks at the cluster border, and therefore that $\omega <1$, but instead of trying to guess the exponent $\omega$ we will leave it as an adjustable parameter.

The general term in the master equation, given the hypotheses above, and with $N$ denoting the total number of disks, while $c_a$ and $c_d$ being arbitrary dimensionless constants associated with attachment and detachment events, reads
\beq \label{ME}
\begin{split}
\frac{dn_k}{dt} = & \frac{c_d}{\tau}(S_{k+1}n_{k+1} - S_kn_k) \\
& + \frac{c_a v_0 a}{L^2} n_1 (S_{k-1}n_{k-1} - S_kn_k), \\
\end{split}
\eeq
where $k=2,\ldots,N-1$. The equations for $k=N$ and $k=1$ are special. For the former, one has
\beq
\frac{dn_N}{dt} = -\frac{c_d}{\tau}S_Nn_N + \frac{c_av_0 a}{L^2} n_1 S_{N-1}n_{N-1},
\eeq
whereas for the latter,
\beq
\begin{split}
&\frac{dn_1}{dt} = \frac{c_d}{\tau}(2S_2n_2+S_3n_3+\ldots+S_Nn_N) - \\ &\frac{c_av_0 a}{L^2} n_1 (2S_1n_1+S_2n_2+\ldots+S_{N-1}n_{N-1}).
\end{split}
\eeq
This set of equations conserve the total number of disks, $\sum_{k=1}^N k\,n_k(t)=N$. Using $\tau$ as a unit of time and absorbing a factor $\pi$ from $\phi$ in the  constant $c_a$, Eq.~\eqref{ME} becomes
\beq\label{ME1}
\begin{split}
\frac{d n_k}{dt} &= c_d\, (S_{k+1} n_{k+1}-S_k n_k) + \\ &\left(\frac{c_a\phi\mathrm{Pe}}{N}\right) \,n_1 \, (S_{k-1} n_{k-1}-S_k n_k).
\end{split}
\eeq
Although the dependence on $\phi$ and $\mathrm{Pe}$ is explicit, it should be kept in mind the possibility of the exponent $\omega$ in $S_k$, as well as the constants $c_a$ and $c_d$ to depend on $D_T$.

\subsubsection{The stationary solution}

It can be checked that the following exponential times power-law dependence,
\beq\label{statCSD}
n_k = N\frac{c_d}{c_a\phi\Pe} e^{\alpha k}\,k^{-\omega},
\eeq
is the stationary solution of Eq.~\eqref{ME1} for any $\alpha$, which will be taken as an adjustable parameter. The factor $c_d/c_a$ is implicitly dependent on $\alpha$ and $\omega$ through the normalization condition,
\beq\label{normCSD}
\sum_{k=1}^N e^{\alpha k}\,k^{1-\omega} = \frac{c_a\phi\Pe}{c_d}.
\eeq

When $\alpha < 0$, the cluster size distribution $n_k$ decreases monotonically with cluster size $k$, indicating the absence of macroscopic aggregation. In contrast, for $\alpha > 0$, $n_k$ develops a local maximum at $k = N$, signaling the emergence of a dominant cluster. 

\begin{figure}[h]
\centering
\includegraphics[width=0.4\textwidth]{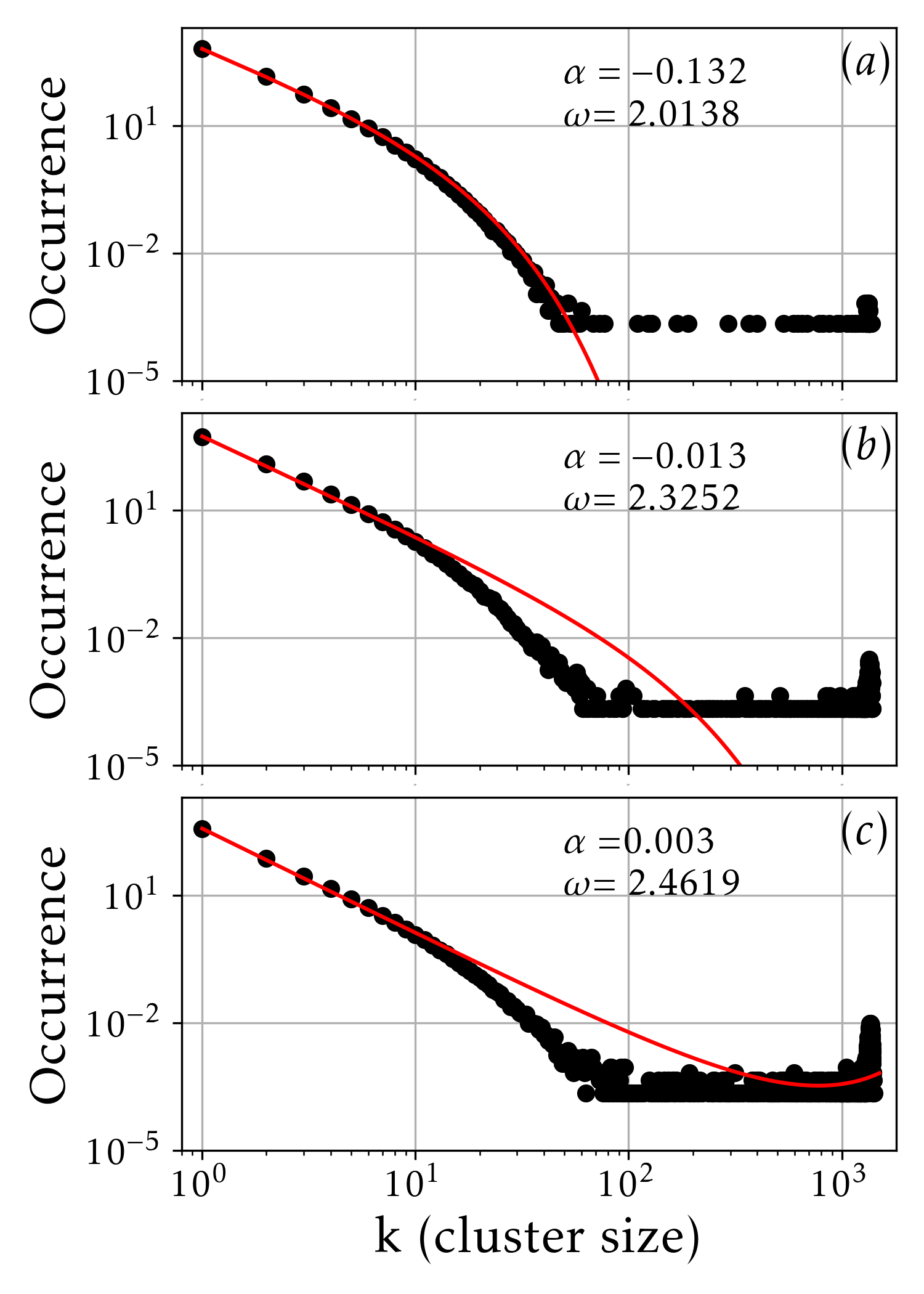}
\caption{\label{fig:CSDFit} The average over 4500 samples of the number of clusters of a given size present (black symbols) and the master equation prediction, $n_k \propto e^{\alpha k}k^{-\omega}$ (red line). The system considered consisted in 1500 disks, with filling fractions $\phi=\{0.13, 0.14, 0.15\}$ (top to bottom), ${\rm Pe}=10^3$, and $D_T=50\,a^2/\tau$. The transition happens when $\alpha=0$ (purely power-law behavior). When $\alpha>0$ the micro-clustering state is \emph{globally} unstable.}
\end{figure}

From Eq.~\eqref{normCSD}, the value of $c_a\phi\Pe/c_d$ that leads to a pure power-law ($\alpha=0$) is
\beq
\left(\frac{c_a\phi\Pe}{c_d}\right)_c = \sum_{k=1}^N k^{1-\omega} \sim N^{2-\omega}.
\eeq
Since one observes clustering with finite values of $\phi \,\mathrm{Pe}$, and in systems with very large $N$, this implies that $\omega \gtrsim 2$, something verified in the fittings shown in Figure \ref{fig:CSDFit}. 

Values of $\omega > 1$ contradict the intuitive expectation that $S_k \propto k^\omega$ should approximate the number of disks at the perimeter of a $k$-cluster. This expectation implicitly assumes that attachment and detachment events occur independently. The observed deviation suggests the presence of correlations in these processes, e.g., the attachment or detachment of dimers and trimers, as illustrated in the video provided in the Supporting Information~\cite{supplementary_video}. These correlations are effectively captured in our single-particle model for the CSD via the parameter $\omega$.

When analyzing the frequency of occurrences for clusters of different sizes across multiple snapshots, we observed that the distribution of occurrences closely resembles a binomial distribution. We therefore fitted the model CSD, Eqs.~\eqref{statCSD} and \eqref{normCSD}, to the numerically averaged data using a maximum likelihood estimator based on the binomial distribution (instead of Gaussian); see Appendix~\ref{fitting_si} for details. Figure~\ref{fig:CSDFit} illustrates representative examples of these fits for three values of $\phi$, with the values of $\alpha$ and $\omega$ obtained. It shows the onset of macro-clustering, evidenced by the emergence of a peak in the CSD at large cluster sizes.

\begin{figure}[h]
        \centering
        \includegraphics[width=0.45\textwidth]{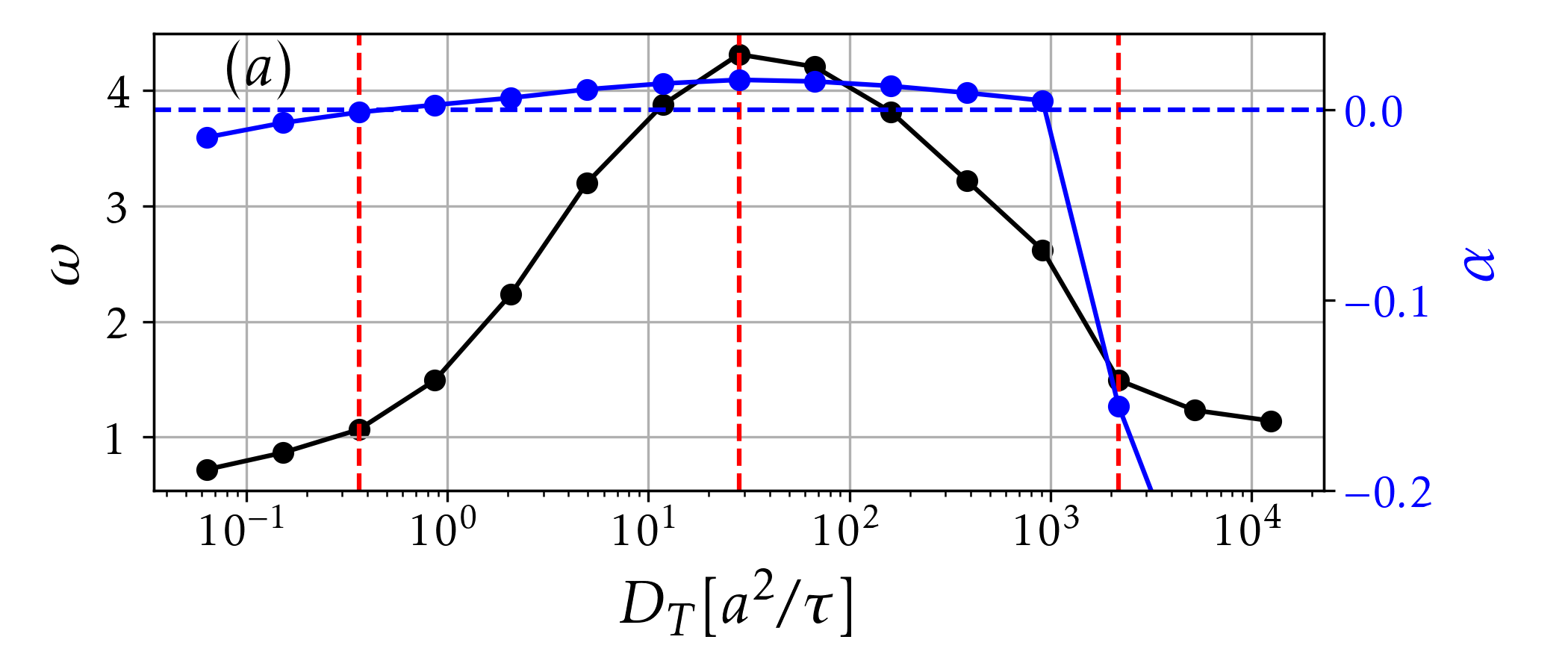}
        
    \includegraphics[width=0.45\textwidth]{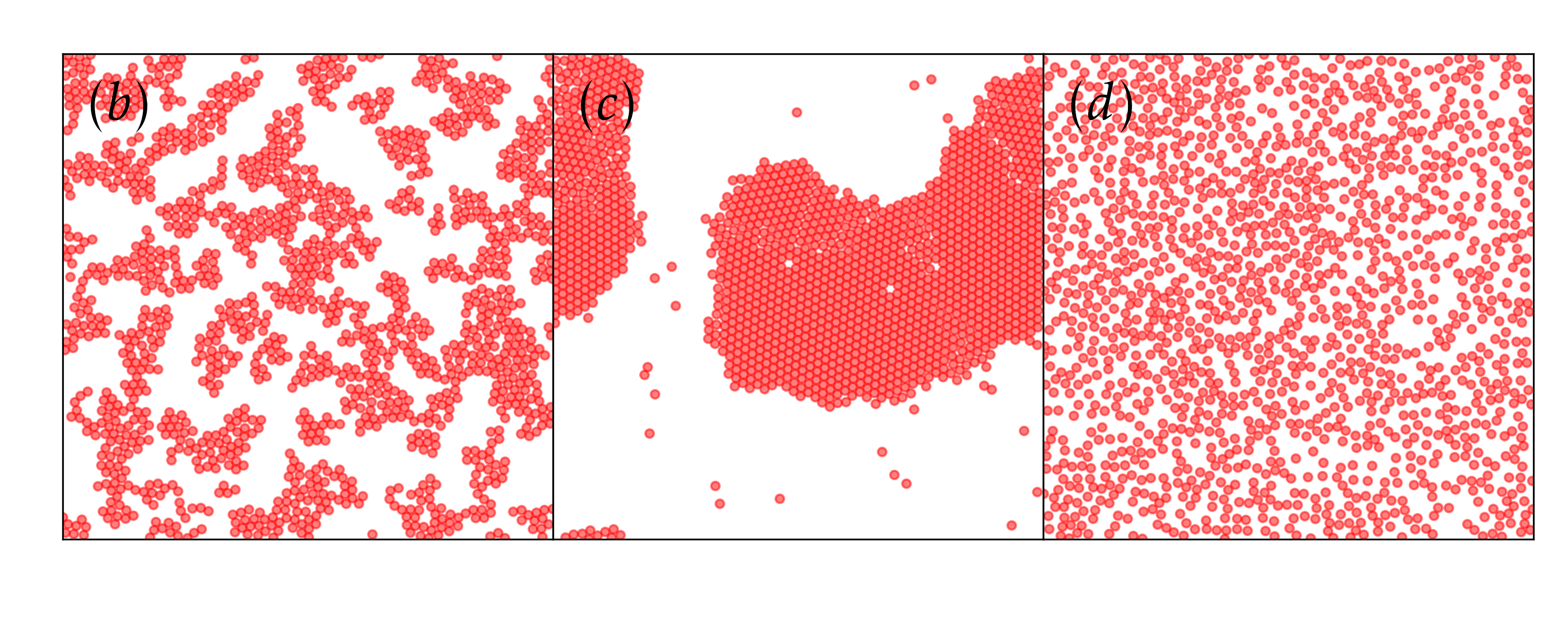}
        \caption{\label{dtw}(a) Fitted values of $\omega$ (black, left axis) and $\alpha$ (blue, right axis) for different values of $D_T$, $\phi = 0.35$, $N= 1000$ and ${\rm Pe} = 10^3$. The vertical dashed red lines correspond to the three snapshots shown in (b-d), with $D_T\in \{0.36,\,28,\,2.2\times10^3\}\,[a^2/\tau]$, respectively.}
    \end{figure}

Figure~\ref{dtw}(a) illustrates the effect of varying translational diffusivity $D_T$, at fixed $\phi$ and Pe, on the CSD parameters $\alpha$ and $\omega$. The parameter $\alpha$ (blue curve) is positive within the range $D_T \in [0.36, 1000]\,[a^2/\tau]$, indicating the occurrence of MIPS. This positive region reflects the reentrant behavior induced by changes in $D_T$. Correspondingly, the parameter $\omega$ peaks within the same interval, presumably due to MIPS-enhanced correlations in particle attachment and detachment events. Panels (b)--(d) in Fig.~\ref{dtw} provide representative snapshots illustrating this $D_T$-driven reentrant scenario.

%\FC{We note that the non-monotonic behavior of $f_{\rm max}$ with respect to $D_T$ was again observed, now reflected in the master equation parameters. In particular, both $\alpha$ and $\omega$ exhibited peaks at the same value of $D_T$, marking the transition between regimes where $\alpha<0$ and $\alpha>0$, as shown in Fig.~\ref{dtw}. The peak occurred at $D_T = 28\,[a^2/\tau]$, corresponding to $f_{\rm max}\sim1$, which indicated the presence of genuine MIPS driven by translational noise (Figs.~\ref{dtw}(a,c)). However, both Fig.~\ref{dtw}(b) and Fig.~\ref{dtw}(d) displayed $\alpha<0$ and similar values of $\omega$, despite showing clearly distinct macroscopic configurations. This outcome was consistent with the expected behavior for $\alpha$ close to zero, where the cluster size distribution became predominantly power law in systems that had not undergone evaporation (Fig.~\ref{dtw}(b)).}

\subsection{Hydrodynamic approach from particle simulations}\label{Dens}

As shown in Ref.~\cite{Bialk__2013}, the mean-field single particle probability density function, $\psi_1(\vr,\theta,t)$, of an ensemble of interacting particles executing active Brownian motion~\cite{Stenhammar2,Marchetti12,RojasVega2023} obeys the Fokker-Planck equation
\begin{equation}
    \partial_t\psi_1  = - \nabla \cdot \left[v \hat{\boldsymbol{e}}_\theta - D\nabla\right]\psi_1 +\frac{1}{2\tau}\partial_{\theta \theta}\psi_1,
\label{hd1}
\end{equation}
where the self-propulsion speed and translational diffusivity, $v$ and $D$, respectively, depend on the \emph{local} filling fraction $\varphi$. Comparing this expression with the non-interacting case, Eq.~\eqref{FP}, one sees that the effect of the interactions is to renormalize the bare $v_0$ and $D_T$.

\begin{figure}[htb]
\centering
\includegraphics[width=0.25\textwidth]{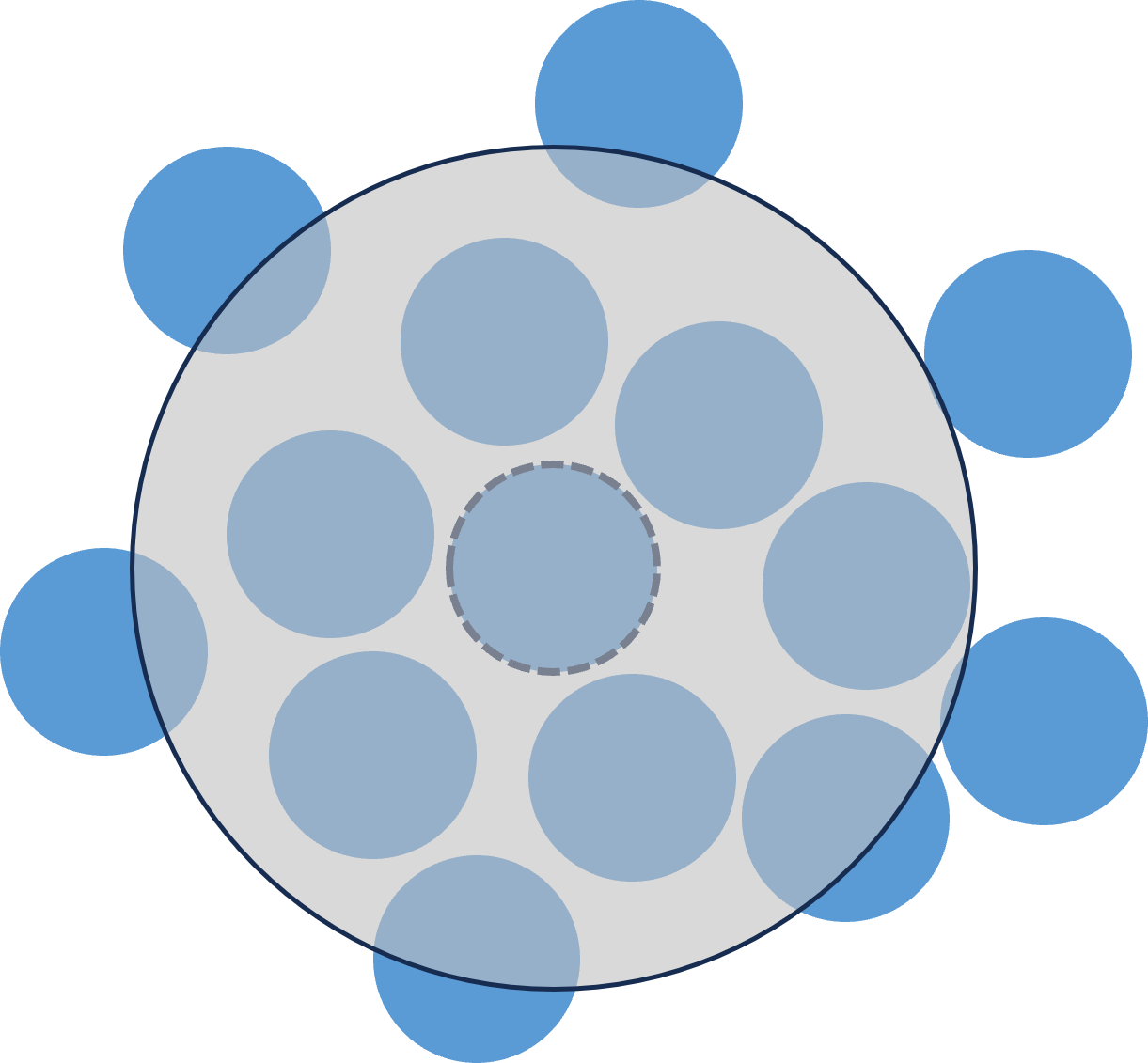}
\caption{\label{fig:SearchDisk} Illustration of the method used to obtain the local filling fraction. A circular area of radius $2.4\,a$ is drawn centered around each individual disk. The fraction of this area containing disks is the local filling fraction of the central disk. It follows that $\varphi_{\rm min}\approx 0.1736$ and $\varphi_{\rm max}\approx 0.9069$.}
\end{figure}

Rather than modeling $v(\varphi)$ and $D(\varphi)$, as done in Refs.~\cite{Bialk__2013,Cates15}, we directly extracted these quantities from the simulation.
We assigned each disk to a bin based on its $\varphi$, see Fig.~\ref{fig:SearchDisk}. 

\begin{figure}[!h]
\hspace{-0.5cm}
\includegraphics[width=0.5\textwidth]{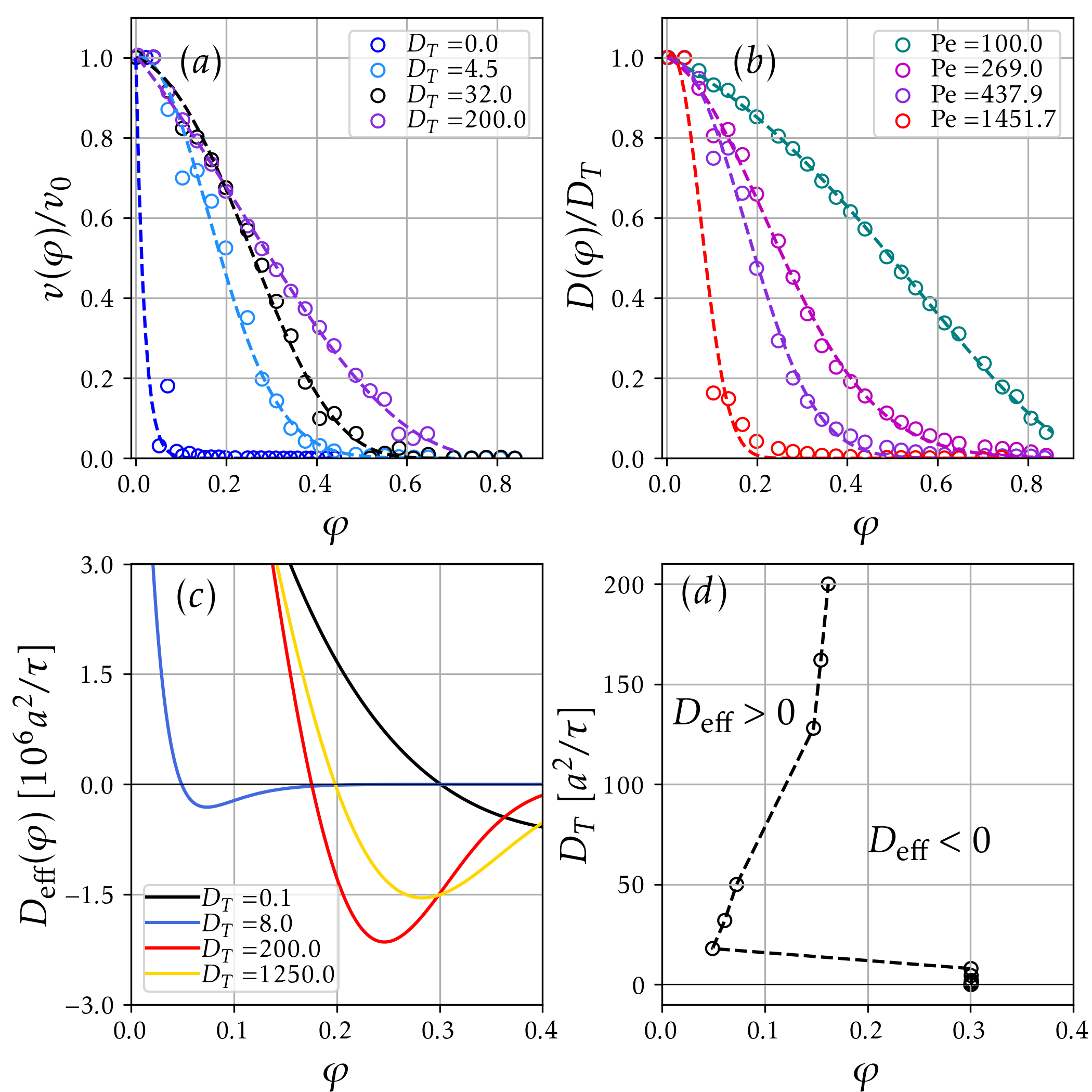}

%\centering
%\includegraphics[width=0.45\textwidth]{extrahydro.png}
\caption{\label{fig:LocalPhi}(a) Self-propulsion speed as a function of the local filling fraction for ${\rm Pe}= 7.74\times10^2$ and different values of $D_T$ (in units of $a^2/\tau$), (b) translational diffusion coefficient as a function of the local filling fraction for $D_T=0.5\,a^2/\tau$ and different values of Pe, (c) effective diffusion coefficient as a function of the local filling fraction, obtained from $v(\varphi)$ and $D(\varphi)$ using Eq.~\eqref{Deff}, for ${\rm Pe}=3.48\times10^3$ and different values of $D_T$ (in units of $a^2/\tau$), (d) the roots of $D_{\rm eff}(\varphi)$ for ${\rm Pe}=3.48\times10^3$ and different values of $D_T$. The dashed line is the spinodal line. It separates the region where the micro-clustering state is locally stable ($D_{\rm eff}>0$) from the region where it is locally unstable ($D_{\rm eff}<0$). For all panels, $N = 1500$ and $\phi=0.196$.}
\end{figure}

For each bin, we computed the mean displacement along the propulsion direction and the displacement variance over a single time step $dt$, and used ${v(\varphi)dt=\langle d\vr\cdot\vee_\theta \rangle}$ and $4D(\varphi)dt = \langle d\vr^2 \rangle - \langle d\vr\cdot\vee_\theta \rangle^2$. This procedure was repeated across multiple snapshots to improve statistical accuracy. A representative result is shown in panels (a) and (b) of Fig.~\ref{fig:LocalPhi}. Both $v$ and $D$ decay monotonically with $\varphi$ as expected, but with significant deviations from the linear behavior that is sometimes assumed~\cite{Bialk__2013}. 

As also shown in Ref.~\cite{Bialk__2013}, from the Fokker-Planck Eq.~\eqref{hd1} one may derive, in the hydrodynamic approximation, a nonlinear diffusion equation for the local density $\rho(\vr,t)$. In terms of the local filling fraction, $\varphi(\vr,t)=(\pi a^2) \rho(\vr,t)$ it reads:
\beq
\partial_t \varphi = \nabla\cdot [D_{\rm eff} \nabla\varphi], 
\eeq
with,
\beq
D_{\rm eff} (\varphi) \equiv  \tau v \frac{dv}{d\varphi}\varphi + \tau v^2 +D. \label{Deff}
\eeq
Figure \ref{fig:LocalPhi}(c)
shows examples of $D_{\rm eff}(\varphi)$ obtained with $v(\varphi)$ and $D(\varphi)$ extracted from the simulation as explained above. 

The \emph{local} instability of the micro-clustering state is indicated by $D_{\rm eff} < 0$, i.e., negative effective diffusivity amplifies infinitesimal density fluctuations among the disks. A point on the spinodal surface is defined by the triplet $(v_0, D_T, \phi = \varphi^*)$, where $\varphi^*$ is the smallest positive root of $D_{\rm eff}(\varphi)$ corresponding to the pair $(v_0, D_T)$. The approximate spinodal lines derived in this way are shown in Figure \ref{fig:LocalPhi}(d) in the $D_T$-$\phi$ plane, and in Figure \ref{fig:PhsDiag} in the Pe-$\phi$ plane. In the latter case, the overall agreement between the region of large $f_{\rm max}$ and the 
domain of local instability is satisfactory, especially considering that $f_{\rm max}$
is a relatively crude order parameter. Furthermore, this analysis shows that the non-monotonic trends obtained from $f_{\rm max}$ are also obtained from the spinodal line via the hydrodynamic theory.

\section{Conclusion}
\label{conclusion}
Translational noise, when added to persistent motion, plays a distinctive role in systems of hard particles. Specifically, it provides the only mechanism for displacements orthogonal to the persistent velocity, as first mentioned in Ref.~\cite{Ber14}. Without translational noise, the clustering of hard active particles is filamentous, with no macroscopic cluster formation. For intermediate values, the added degree of freedom introduced by independent displacements favors the emergence of larger, rounder, and more compact clusters. However, we observed that when the noise amplitude is too high, individual particle motion becomes predominantly diffusive, and clustering is suppressed. 

We refined the master equation model proposed in Ref.~\cite{Ber14} by introducing a parameter associated with the effective cluster perimeter. We derived an expression for the stationary cluster size distribution and identified a clear marker of \emph{global} clustering instability. The numerical fit of the cluster size distribution to the theoretical formula was satisfactory and provided valuable insights. Specifically, the values obtained for the exponent $\omega$ of the effective perimeter parameter indicate that the disks' attachment to and detachment from clusters are correlated. Observations of the dynamics confirm this correlation, that is, disks frequently detach in groups rather than individually. See video in the Supporting Information~\cite{supplementary_video}. 

We developed an algorithm to extract the effective local self-propulsion speed and translational diffusion coefficient of a disk from numerical simulations, explicitly accounting for the influence of its local environment. The algorithm operates on a single time step to ensure that the local environment is well-defined. The resulting dependence of $v$ and $D$ on the local filling fraction was then incorporated into the hydrodynamic equations from Ref.~\cite{Bialk__2013} to determine the critical parameters for the \emph{local} instability of the micro-clustering state, i.e., the spinodal surface.

In conclusion, our findings open new perspectives on the role of translational noise in Active Matter. In particular, lattice-based models of active matter with effectively hard interactions may serve as a promising test bed to further explore the effects of translational noise and validate the results presented here. Other directions may include analyzing the impact of translational noise in different spatial dimensions, lattices, or in systems where strong attractive forces are present -- conditions that, in passive systems, are already known to produce gel-like behavior. In such scenarios, the critical value of $D_{T}$ required to suppress clustering might increase, potentially enhancing the phenomena observed in our study.

\acknowledgements

F.H.\ thanks CAPES for financial support, the Center for Computing in Engineering and Sciences at Unicamap and the HPC Cluster Coaraci, made available under FAPESP grants 2013/08293-7 and 2019/17874-0, respectively, the Brazilian LNCC (Laboratório Nacional de Computação Científica), for access to the SDumont Cluster, under the SINAPAD/2014 project 01.14.192.00. P.d.C.\ was supported by Scholarships No.\ 2021/10139-2 and No.\ 2022/13872-5 and ICTP-SAIFR Grant No.\ 2021/14335-0, all granted by São Paulo Research Foundation (FAPESP), Brazil.

\section*{DATA AVAILABILITY}

Data will be made available upon request.

\bibliography{References}

\appendix

\onecolumngrid

\section{Mean-squared displacement of the individual dynamics}
\label{app:msd}

The velocity self-correlation function at long times implied by Eq.~\eqref{Langevin} is~\cite{Marchetti12}:
 \beq
 \langle \vv(t)\cdot \vv(s) \rangle \sim v_0^2 e^{-|t-s|/2\tau},
 \eeq
 showing the persistence of the direction of motion for time intervals of the order of the persistence time $\tau=1/2D_R$. 

\begin{figure}[h]
\centering
\includegraphics[width=0.45\textwidth]{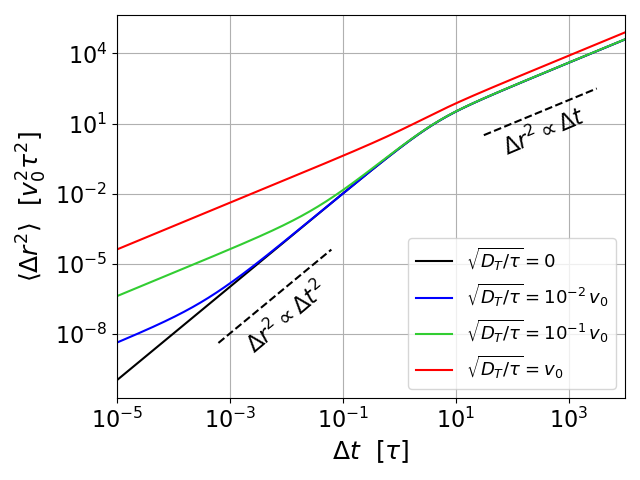}
\caption{\label{fig:Furth} The mean-squared displacement of an isolated disk, Eq.~\eqref{MSD}. In the absence of translational noise ($D_T=0$, black curve), the MSD follows the behavior of a persistent random walk, as described by Furth's formula, transitioning from ballistic motion at short times to diffusive motion at longer times. When the diffusive root mean-squared displacement becomes larger than the persistence length ($v_0\tau$), or equivalently, when $\sqrt{D_T/\tau} > v_0$, the ballistic regime is effectively suppressed, leaving the dynamics dominated by diffusion.}
\end{figure}

The mean-squared displacement (MSD) is
\beq\label{MSD}
\begin{split}
\frac{\langle  \Delta\vr\cdot\Delta\vr \rangle  }{(v_0\tau)^2} &\sim 4\left( \frac{\Delta t}{\tau} -2 + 2\,e^{-\Delta t/2\tau} \right) \\ &+ 4\left( \frac{D_T}{v_0^2 \tau} \right) \frac{\Delta t}{\tau},
\end{split}
\eeq
where $\Delta \vr\equiv \vr(t+\Delta t) - \vr(t)$ and $t\gg\tau$.
The first term, known as Furth's formula~\cite{Fur20}, exhibits a transition from ballistic to diffusive behavior and is the result of the persistent random-walk ingredient. The second term corresponds to a purely diffusive behavior due to the translational noise. The MSD versus time is displayed in Fig.~\ref{fig:Furth}.

\section{On the fitting of the CSDs}
\label{fitting_si}

We found that the distribution of cluster occurrences of a given size at different times follows a Binomial distribution. This result has practical implications for fitting cluster distribution data. 

To support this claim, we report here the occurrences of clusters of different sizes across $N_s=500$ sample snapshots. The system considered used as parameters: $N=2500$ (the number of disks), $\phi=0.13$, ${\rm Pe}=1000$ and $D_T=50\,a^2/\tau$.

\begin{figure}[!h]
\centering
\includegraphics[width=0.45\textwidth]{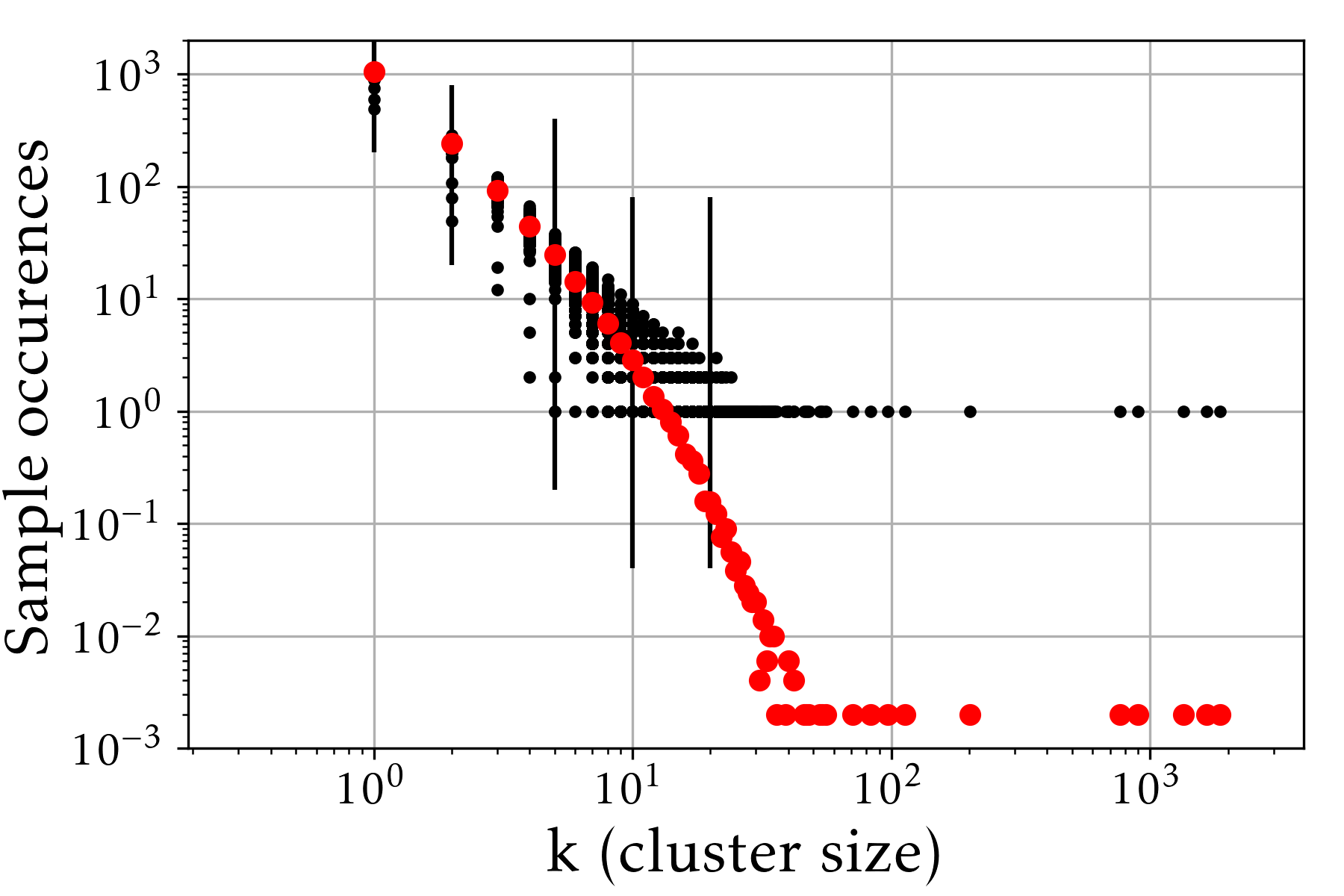}
\caption{\label{fig:SI_rawdata}The set of sample occurrences, $\{n_k^s\},~(s=1,...,N_s)$ against cluster size $k$. Each non-zero occurrence is displayed as a black dot, zero occurrences are omitted. The red symbols are the sample averages, $\langle n_k \rangle \equiv \tfrac{1}{N_s} \sum_{s=1}^{N_s} n_k^s$, where $N_s$ is the number of snapshots. The vertical lines mark the cluster sizes whose occurrences histogram are shown in Fig.~\ref{fig:SI_BinomialDist}. }
\end{figure}

For each of the $k$-clusters $(k=1,...,N)$ we recorded the set of sample occurrences, $\{n_k^s\}$ $(s=1,...,N_s)$.
This raw data is displayed in Fig.~\ref{fig:SI_rawdata}. 

In order to fit the sample averages $\langle n_k \rangle$ (the red symbols in Fig.~\ref{fig:SI_rawdata}) to the stationary solution of the master equation model, Eq.~\eqref{statCSD}, we must find how the sample occurrences are distributed around their average.

\begin{figure*}[!h]
\centering
\includegraphics[width=0.4\textwidth]{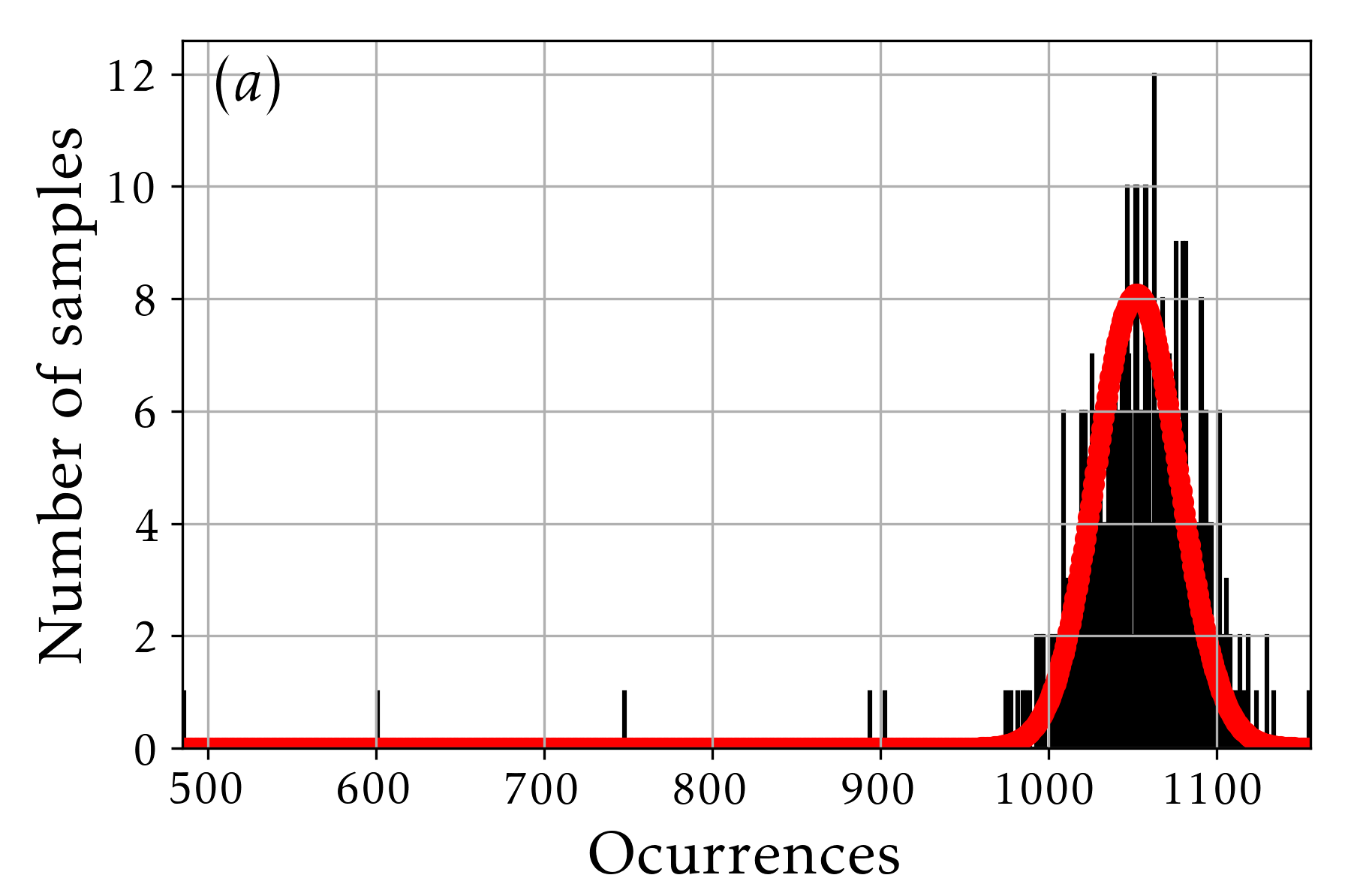}
\includegraphics[width=0.4\textwidth]{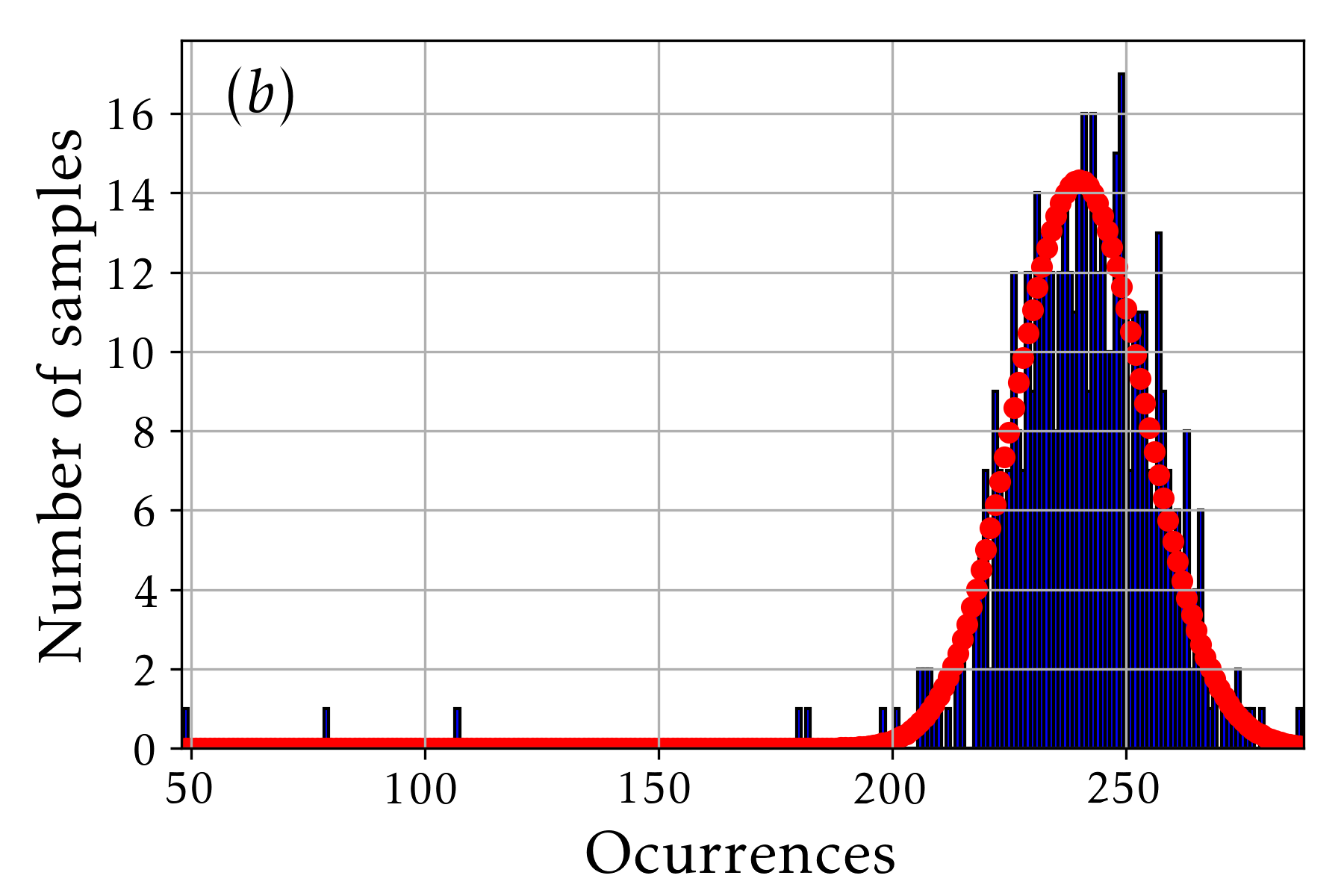}
\includegraphics[width=0.4\textwidth]{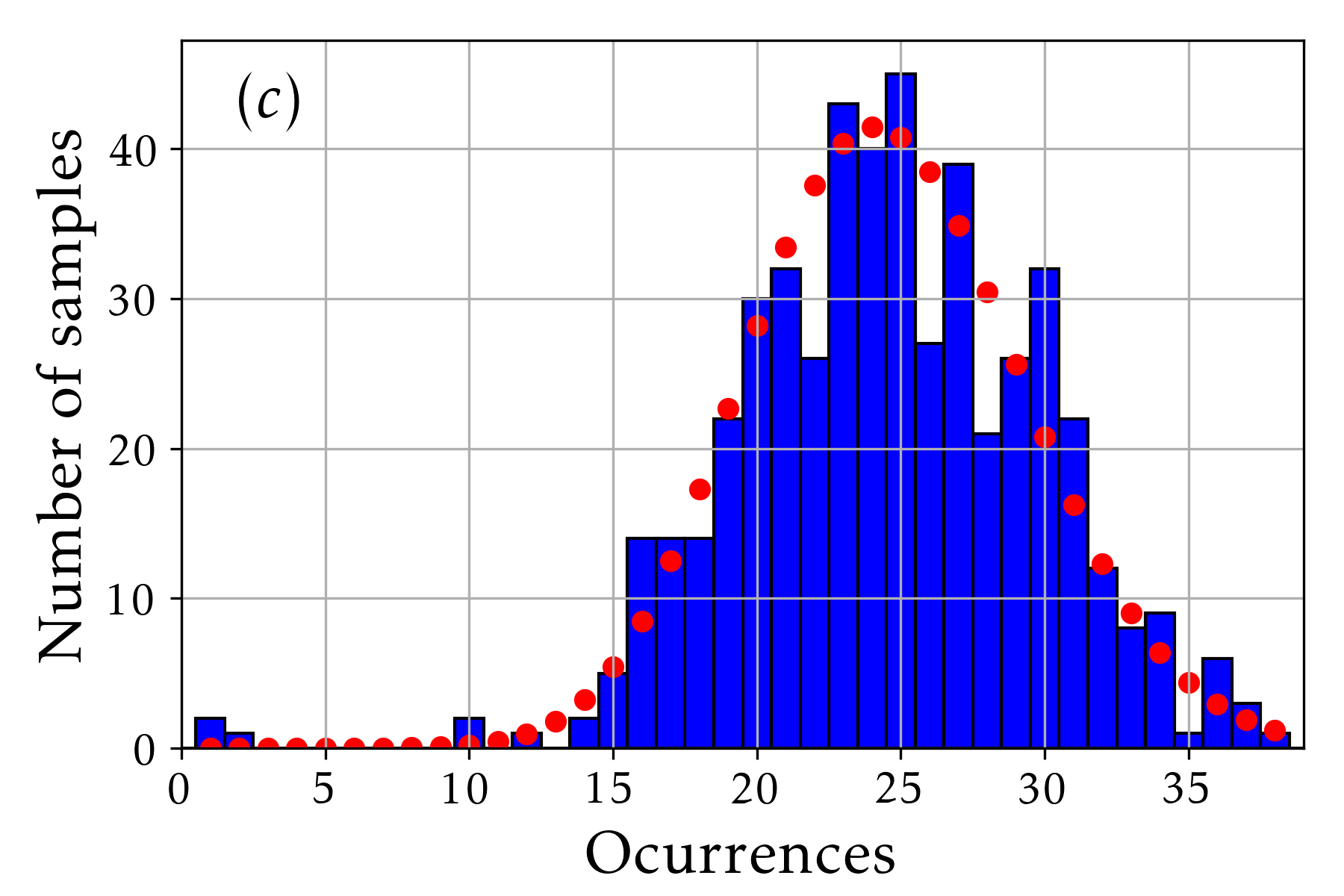}
\includegraphics[width=0.4\textwidth]{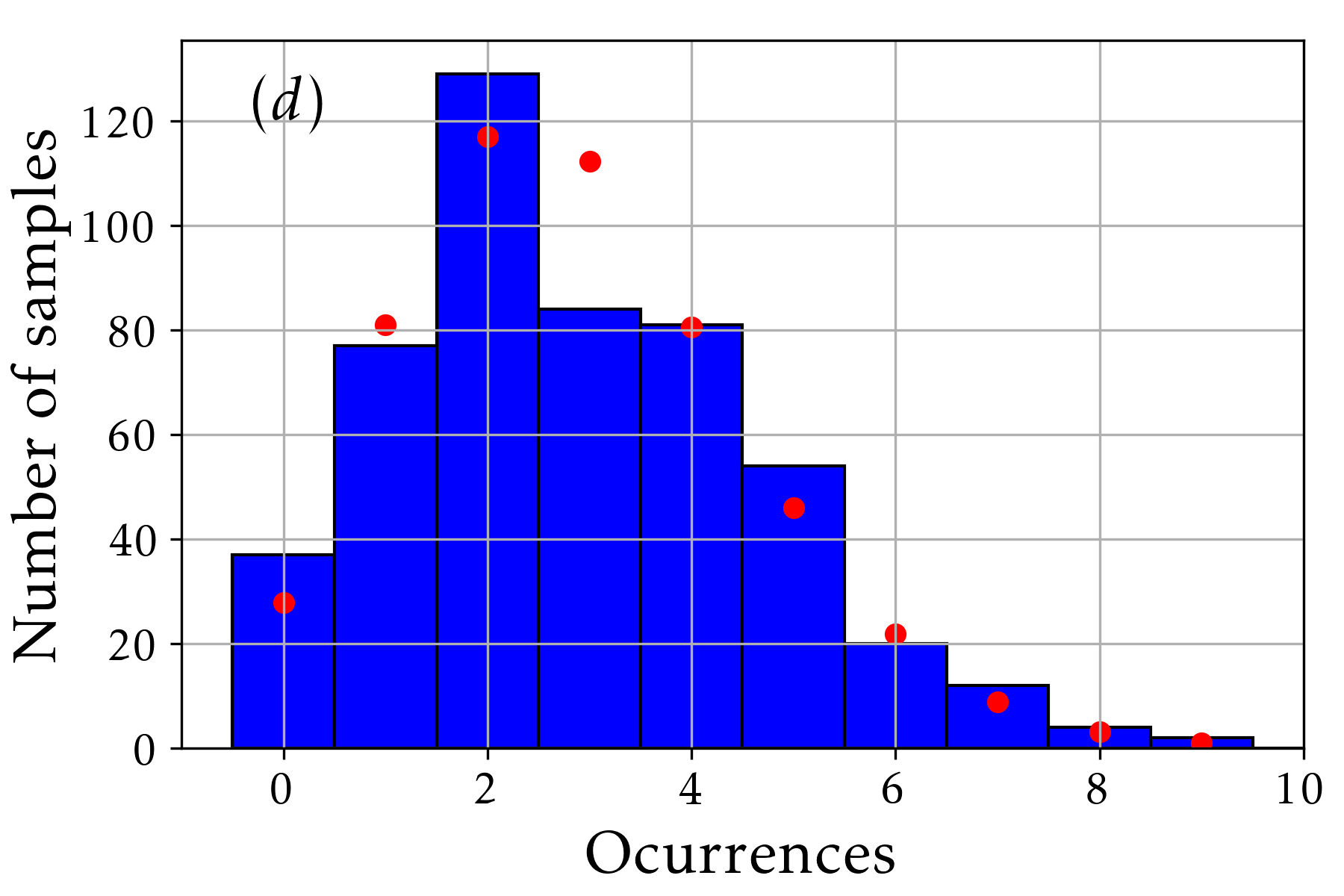}
\includegraphics[width=0.4\textwidth]{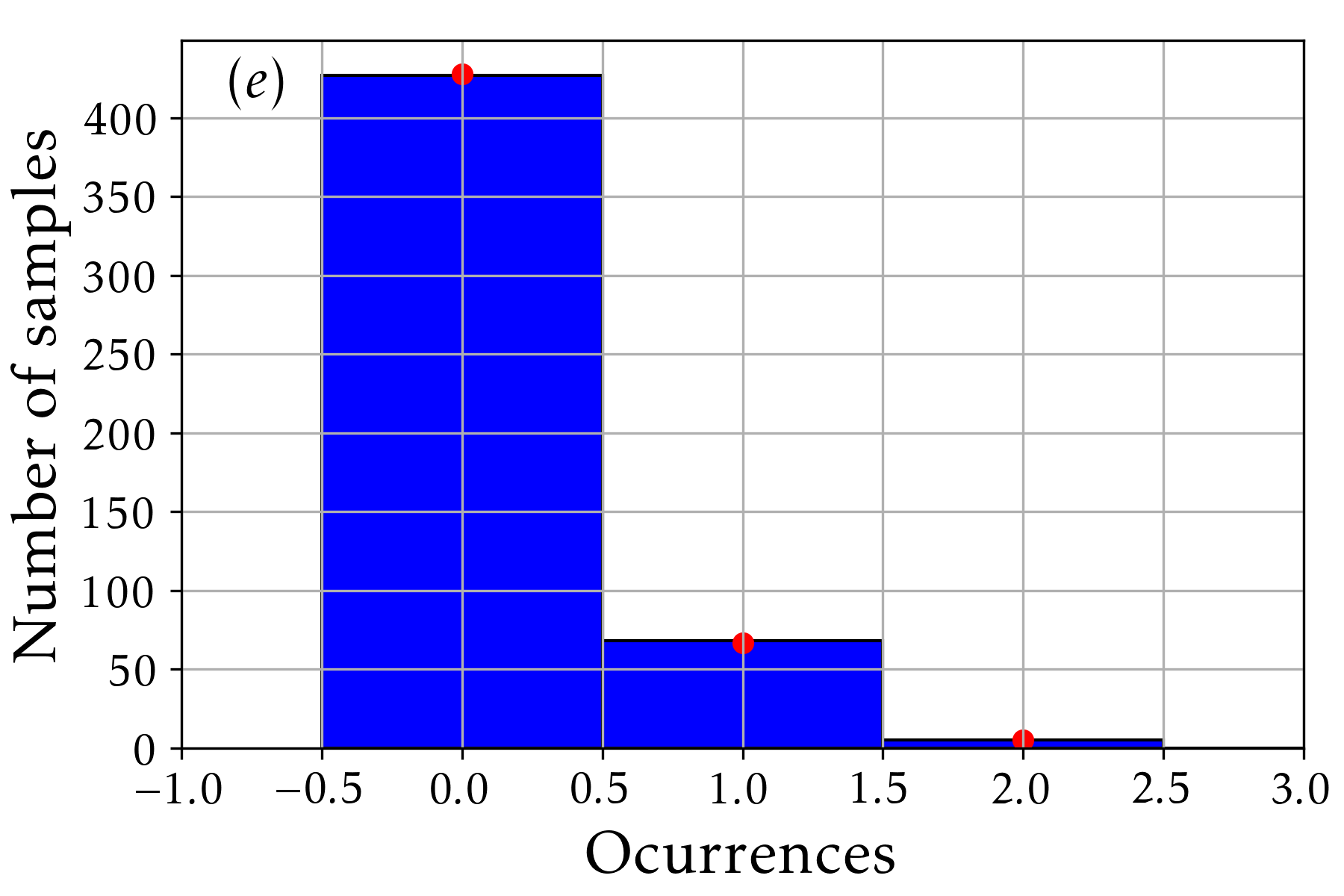}
\caption{\label{fig:SI_BinomialDist} The histogram of the number of samples where a given sample occurrence $n_k\in\{0, 1,..., \lfloor N/k \rfloor\}$ was observed. Panels (a) to (e) show the cases $k=1$, 2, 5, 10 and 20 disks respectively. The number of disks is $N=2500$ and the number of samples is $N_s=500$. The red dots represent the Binomial distribution \eqref{BinomialDist} multiplied by $N_s$.}
\end{figure*}

In Fig.~\ref{fig:SI_BinomialDist} we plot such sample occurrences, i.e., the number of samples where a given value of $n_k\in\{0, 1,..., N_k\equiv\lfloor N/k \rfloor\}$ was observed, for clusters with $k=1$, 2, 5, 10 and 20 disks. The red dots correspond to the binomial probability,
\begin{equation}\label{BinomialDist}
    B(n_k)= \binom{N_k}{n_k} p_k^{n_k}(1-p_k)^{N_k -n_k},
\end{equation}
multiplied by $N_s$, and
with $N_k p_k$ chosen to be equal to the sample average, $\langle n_k \rangle \equiv \tfrac{1}{N_s} \sum_{s=1}^{N_s} n_k^s$.

The probability of the observed full set of occurrences is therefore:
\begin{equation}
    P(\{n_k^s\}) = \prod_{k=1}^N \prod_{s=1}^{N_s} B(n_k^s).
\end{equation}
After replacing the sample average in the Binomial probability \eqref{BinomialDist} by the two parameter master equation prediction, Eq.~\eqref{statCSD}, this becomes the proper maximum likelihood estimator for the fitting.

The function we actually maximized was (after removing all terms independent of the adjustable parameters):
\begin{equation}
\begin{split}\label{F}
    F(\alpha,\omega)=&\sum_{k=1}^N \langle n_k \rangle \log(n_k^{\rm ME})+ \\
    &(N_k-\langle n_k \rangle) \log(N_k- n_k^{\rm ME} ),
\end{split}
\end{equation}
where $N_k=\lfloor N/k \rfloor$, $\langle n_k \rangle \equiv \tfrac{1}{N_s} \sum_{s=1}^{N_s} n_k^s$ is the sample average, and $n_k^{\rm ME}(\alpha,\omega)$ is the corresponding master equation prediction, Eq.~\eqref{statCSD}.

\newpage

%\section{Additional snapshots}
%\label{visualizations}

%\FC{To illustrate the correlation between $f_{\rm max}$ and the sign of $D_{\rm eff}$, Fig.~\ref{fig:SPINODAL} depicts snapshots corresponding to $\phi =0.1$ in Fig.~\ref{fig:LocalPhi}(d).}
%\begin{figure*}[h]
 %   \centering
%\includegraphics[width=0.55\textwidth]{figure_AC.png}
%\caption{\label{fig:SPINODAL}\FC{Snapshots for Pe = $3480$, $\phi = 0.1$, $N=986$, and, in units of $a^2/\tau$,  (a) $D_T = 5$ ($D_{\rm eff} >0$), (b) $D_T = 28$ ($D_{\rm eff} <0$) and (c) $D_T = 200$ ($D_{\rm eff} >0)$. The value of $f_{\rm max}$ is shown in the top left of each panel.}}
%\end{figure*}

%\section{Variation of $\omega$ according to $D_T$}
%\label{varw}

%\FC{We found that the behavior of the exponent $\omega$ displays non-monotonic behavior with the translational diffusivity, $D_T$. This is shown in Fig.~\ref{fig:OMEGADT}, where we observe that $\omega$ has a peak at true MIPS.}
%\begin{figure}[h]
 %   \centering
  %  \includegraphics[width=0.4\textwidth]{fig_appendixCA.png}
  %  \includegraphics[width=0.7\textwidth]{appendix_Dt_snaps.png}
  %  \caption{\label{fig:OMEGADT}\FC{(a) Values of $\omega$ obtained via fitting the CSDs for ${\rm Pe} = 10^3, \phi = 0.35$ and varying $D_T$. The dashed vertical lines indicate the values of $D_T$ used in the snapshots shown in (b) $D_T = 0.5\,[a^2/\tau]$, (c) $D_T = 12.5\,[a^2/\tau]$ and (d)  $D_T = 2.5 \times 10^3 [a^2/\tau]$. CSDs were collected for a system of $N=1000$ disks, whereas the snapshots used $N=3.5\times10^3$ disks.}}
%\end{figure}

\end{document}